\documentclass[aps,prd,preprintnumbers,showpacs,showkeys,nofootinbib,
superscriptaddress,fleqn,floatfix,tightenlines]{revtex4-1}
\usepackage{amsmath,amsfonts,amssymb,amscd,amsxtra,amsthm}
\usepackage{graphicx}  
\usepackage{epstopdf}
\usepackage{dcolumn}  
\usepackage{bm}          
\usepackage{slashed}
\usepackage[utf8]{inputenc}

\usepackage[normalem]{ulem} 
\usepackage[dvipsnames]{xcolor} 
\renewcommand\sout{\bgroup \color{red} \ULdepth=-.5ex \ULset}

\begin{document}
\preprint{INHA-NTG-03/2016}
\title{Parity-violating $\pi NN$ coupling constant from the
  flavor-conserving effective weak chiral Lagrangian}

\author{Chang Ho Hyun}
\email{hch@daegu.ac.kr}
\affiliation{Department of Physics Education,
Daegu University, Gyeongsan 38453, Korea}

\author{Hyun-Chul Kim}
\email{hchkim@inha.ac.kr}
\affiliation{Department of Physics,
Inha University, Incheon 22212, Korea}
\affiliation{School of Physics, Korea Institute for Advanced Study
  (KIAS), Seoul 130-722, Republic of Korea}

\author{Hee-Jung Lee}
\email{hjl@chungbuk.ac.kr}
\affiliation{Department of Physics Education,
Chungbuk National University, Cheongju 28644, Korea}

\date{\today}

\begin{abstract}
We investigate the parity-violating pion-nucleon-nucleon coupling constant
$h^1_{\pi NN}$, based on the chiral quark-soliton model.
We employ an effective weak Hamiltonian that takes into account the
next-to-leading order corrections from QCD to the weak interactions at the
quark level. Using the gradient expansion, we derive the
leading-order effective weak chiral Lagrangian with the low-energy
constants determined. The effective weak chiral Lagrangian is
incorporated in the chiral quark-soliton model to calculate the
parity-violating $\pi NN$ constant $h^1_{\pi NN}$. We obtain a value of
about $10^{-7}$ at the leading order. The corrections from the
next-to-leading order reduce the leading order result by about 20~\%.
\end{abstract}

\maketitle

\section{Introduction}
The electroweak interactions have been tested and confirmed mainly by
parity-violating lepton scattering, decays of hadrons, and
$\beta$ decays of nuclei. Recently, Parity-violating (PV) hadronic
processes play yet another important role of a touchstone to
examine the standard model (SM) and physics beyond the standard model
(BSM)~(See for example recent reviews~\cite{RamseyMusolf:2006dz,
Holstein:2009zzc, Cirigliano:2013xha, Cirigliano:2013lpa,
Haxton:2013aca, Schindler:2013yua}).
There are mainly two different ways of describing PV hadronic
reactions: One is to consider one-boson exchanges such as $\pi$,
$\rho$, and $\omega$ mesons  \`{a} la the strong nucleon–nucleon
($NN$) potential~\cite{Tadic:1969xx, Donoghue:1975xk,
  Desplanques:1979hn}. The other is to employ effective field
theory~\cite{Zhu:2004vw, Schindler:2013yua}. In both
methods, the PV pion-nucleon coupling constant is the most essential
quantity, since it governs the PV hadronic processes in long range .
Desplanques, Donoghue and Holstein (DDH)~\cite{Desplanques:1979hn}
estimated the value of the PV $\pi NN$ coupling constant, also known
as the so-called ``DDH best value'': $h^1_{\pi NN}=4.6 \times
10^{-7}$. A great deal of experimental and theoretical efforts has
been devoted to extract the precise value of the $\pi NN$ coupling
constant (for recent reviews, see \cite{haxton2013, schin2013}). For
example, its contribution is exclusively dominant in the PV asymmetry
in  $\vec{n} p \to d\gamma$~\cite{hyun2001, hyun2005, hyun2007},
and $\vec{n} d \to t \gamma$~\cite{des1986}. The PV $\pi NN$ coupling
constant has been studied in various different theoretical approaches
such as the  Skyrme models~\cite{kaiser1988, weigel1999, ks1993},
quark models~\cite{dubovik1986}, the chiral-quark soliton
model~\cite{Lee:2004tr, Lee:2012fx}, QCD sum rule~\cite{henley1998},
and so on. However, all these values of $h^1_{\pi NN}$ are far from
consensus and are given in the wide range between $10^{-8}$
\cite{kaiser1988} and $\sim 5\times 10^{-7}$ \cite{ks1993}. A recent
analysis of lattice QCD yields $h_{\pi NN}^{1,\mathrm{con}} =
(1.099\pm0.505_{-0.064}^{+0.058})\times 10^{-7}$ for which only the
contribution of the connected diagrams to $h_{\pi NN}^1$ has been
considered~\cite{wasem2012}. On the experimental side, though the
accuracy of the measurements has been much improved, an upper bound on
the value of $h^1_{\pi NN}$~\cite{Gericke:2011zz} is only known. Thus,
more systematic and quantitative studies are required in order to
obtain the value of the PV $\pi NN$ coupling constant.

The main dynamical origin of hadronic parity violation comes from the
flavor-conserving effective weak Hamiltonian, which was already
investigated~\cite{Donoghue:1975xk, Korner:1978sz,
  Guberina:1978wg, Desplanques:1979hn, Karino:1980vn, Karino:1981yj,
   Dai:1991bx}. In particular, the PV $\pi NN$ coupling constant can be
obtained from the isovector ($\Delta I=1$) effective weak Hamiltonian,
which was first derived in Ref.~\cite{Dai:1991bx} at the one-loop
level with the effects of heavy quarks taken into account. Very
recently, Tiburzi~\cite{Tiburzi:2012hx} investigated systematically
the $\Delta S=0$ effective weak Hamiltonian with QCD corrections at
next-to-leading order (NLO). The effects of the NLO corrections have
changed the Wilson coefficients about $(10-20)\,\%$ at the typical scale
of light hadrons ($\mu=1\,\mathrm{GeV}$). Considering the fact that
the PV $\pi NN$ coupling constant is very tiny, we expect that
the corrections from QCD at NLO may come into play. Thus, it is of
great interest to examine the NLO corrections to the PV $\pi NN$
coupling constant.

In the present work, we investigate the PV $\pi NN$ coupling constant,
$h^1_{\pi NN}$, within the framework of the chiral quark-soliton model
($\chi$QSM) together with the effective weak Hamiltonian at
NLO~\cite{Tiburzi:2012hx}. Recently, the present authors computed the
PV $\pi NN$ coupling constant~\cite{Lee:2012fx}
in the same framework, employing the effective weak Hamiltonian from
Ref.~\cite{Desplanques:1979hn}. We first derived the effective weak
chiral Lagrangian, based on the nonlocal chiral-quark model
(N$\chi$QM) from the instanton vacuum associating with the effective
weak Hamiltonian~\cite{Lee:2004tr}. If one performs the gradient
expansion for the effective chiral action of the $\chi$QSM with the
effective weak Hamiltonian, we would obtain exactly the same
expressions starting directly from the effective weak chiral
Lagrangian. Using this gradient expansion, we were able to obtain
the PV $\pi NN$ coupling constant to be about
$1\times 10^{-8}$ at $\mu=1$ GeV. We also found that the $h_{\pi
  NN}^1$ is rather sensitive to the Wilson coefficients. In this
respect, it is of great importance to reexamine the PV $\pi NN$
coupling constant, the $\Delta I=1$ effective weak Hamiltonian being
employed with the NLO QCD effects. As we will show in this work, the
value of $h_{\pi   NN}^1$ indeed turns out to be different from the
previous result. Moreover, the effects from the next-to-leading-order
corrections reduce the reading-order result by about $20~\%$.

The paper is organized in the following order: In Section
\ref{sec:weaklag}, we present briefly the general procedure to obtain
the PV $\pi NN$ coupling constants within the $\chi$QSM. We
first derive the flavor-conserving effective weak chiral Lagrangian,
starting from the nonlocal chiral quark model from the instanton
vacuum. In Section III we compute the correlation function corresponding to the PV
$\pi NN$ coupling constant. In Section \ref{sec:con}, we discuss the
result, and conclude the work.

\section{$\Delta I=1$ Effective weak chiral
  Lagrangian}\label{sec:weaklag}
We start with the $\Delta I=1$ flavor-conserving effective weak
Hamiltonian including the NLO corrections~\cite{Tiburzi:2012hx}, which
is expressed as
\begin{align}
\mathcal{H}_{W}^{\Delta I=1} = \frac{G_F}{\sqrt{2}}
  \frac{\sin^2\theta_W}{3}\sum_{i=1}^{8}c_i(\mu) O_i(\mu)\ ,
\end{align}
where $G_F$ and $\sin\theta_W$ denote the Fermi constant and the
Weinberg angle, respectively. The eight different operators $O_i$ are
defined generically as two-body operators:
$O_i=(\bar{\psi}\gamma_\mu\gamma_5 M \psi) (\bar{\psi} \gamma_\mu N
\psi)$. The $c_i$ stands for the Wilson coefficient corresponding to
the $O_i$, which depends on the renormalization scale
$\mu$. Introducing the Gell-Mann matrices in SU(3) flavor space, we
can write $O_i$ as
\begin{align}
O_1&=\sqrt{\frac{1}{3}} ({\bar \psi}\gamma_\mu\gamma_5\lambda_3\psi)
\bigg({\bar \psi} \gamma_\mu \bigg[\sqrt{2} \lambda_0 +
       \lambda_8\bigg]\psi\bigg)\  , \cr
O_2&= \sqrt{\frac{1}{3}}{\bar \psi}^a \gamma_\mu\gamma_5 \lambda_3\psi^b
{\bar \psi}^b\gamma_\mu \bigg[\sqrt{2} \lambda_0+\lambda_8
     \bigg]\psi^a\ , \cr
O_3& = \sqrt{\frac{1}{3}} ({\bar \psi}\gamma_\mu\lambda_3\psi)
\bigg({\bar \psi}\gamma_\mu \gamma_5 \bigg[\sqrt{2}\lambda_0 +
     \lambda_8\bigg]\psi\bigg)\ , \cr
O_4&=\sqrt{\frac{1}{3}}{\bar \psi}^a\gamma_\mu\lambda_3\psi^b
{\bar \psi}^b\gamma_\mu\gamma_5 \bigg[\sqrt{2} \lambda_0 +
     \lambda_8\bigg]\psi^a\ , \cr
O_5&=\sqrt{\frac{1}{6}}({\bar \psi}\gamma_\mu\gamma_5\lambda_3\psi)
\bigg({\bar \psi}\gamma_\mu \bigg[\lambda_0- \sqrt{2}\lambda_8
     \bigg]\psi\bigg)\ , \cr
O_6&=\sqrt{\frac{1}{6}}{\bar \psi}^a \gamma_\mu \gamma_5 \lambda_3\psi^b
{\bar \psi}^b\gamma_\mu \bigg[\lambda_0-\sqrt{2}
     \lambda_8\bigg]\psi^a\ , \cr
O_7&=\sqrt{\frac{1}{6}}({\bar \psi}\gamma_\mu\lambda_3\psi)
\bigg({\bar \psi}\gamma_\mu \gamma_5 \bigg[\lambda_0 -
     \sqrt{2}\lambda_8\bigg]\psi\bigg)\ ,
\cr
O_8&=\sqrt{\frac{1}{6}}{\bar \psi}^a \gamma_\mu \lambda_3\psi^b
{\bar \psi}^b\gamma_\mu \gamma_5
     \bigg[\lambda_0-\sqrt{2}\lambda_8\bigg]\psi^a\ ,
\label{eq:twobody}
\end{align}
where $\lambda_0$, $\lambda_3$, and $\lambda_8$ are the Gell-Mann
matrices represented in flavor SU(3) space as
$\lambda_0=\sqrt{2/3}\,\mathrm{diag}(1,\,1,\,1)$,
$\lambda_3=\mathrm{diag}(1,\,-1,\,0)$, and
$\lambda_8=\sqrt{1/3}\,\mathrm{diag}(1,\,1,\,-2)$, respectively.
The quark field is given as a triplet in flavor SU(3)
\begin{align}
\psi = \left(\begin{array}{c} u \\ d \\ s \end{array} \right),
\nonumber
\end{align}
where $u$, $d$ and $s$ represent the up, down and strange quark
fields, respectively. The repeated indices $a$ and $b$ designate the
color-singlet contraction and the parentheses $(\bar{\psi}\Gamma \psi)$
without showing the color indices are already color-singlet contracted.
Applying the following Fiertz identity to $O_2, O_4, O_6,$ and $O_8$,
\begin{align}
\delta_{bc}\delta_{ad}=\frac{1}{2}(t^A)_{ab}(t^A)_{cd}
\end{align}
where $t^A$ denote the Gell-Mann matrices in color space,
we are able to express the effective weak Hamiltonian in the following
form
\begin{align}
\mathcal{H}_{W}^{\Delta I=1} &=\frac{G_F}{\sqrt{6}}
                    \frac{\sin^2\theta_W}{3} \bigg\{({\bar
                    \psi}\gamma_\mu \gamma_5\lambda_3\psi)
\bigg({\bar \psi}\gamma_\mu\bigg[\lambda_0 \bigg(\sqrt{2}c_1 +
                    \frac{c_5}{\sqrt{2}} \bigg) + \lambda_8(c_1-c_5)
                    \bigg]\psi\bigg)\cr
&+\frac{1}{2}({\bar \psi}\gamma_\mu\gamma_5 \lambda_3t^A\psi)
\bigg({\bar \psi}\gamma_\mu\bigg[\lambda_0
  \bigg(\sqrt{2}c_2+\frac{c_6}{\sqrt{2}}\bigg)
+\lambda_8(c_2-c_6)\bigg]t^A\psi\bigg) \cr
&+({\bar \psi}\gamma_\mu\lambda_3\psi)
\bigg({\bar  \psi}\gamma_\mu\gamma_5 \bigg[\lambda_0
  \bigg(\sqrt{2}c_3+\frac{c_7}{\sqrt{2}}\bigg)
  +\lambda_8(c_3-c_7)\bigg]\psi\bigg) \cr
&+\frac{1}{2}({\bar \psi}\gamma_\mu\lambda_3t^A\psi)
\bigg({\bar \psi}\gamma_\mu\gamma_5\bigg[\lambda_0 \bigg(\sqrt{2}c_4 +
  \frac{c_8}{\sqrt{2}}\bigg) +\lambda_8(c_4-c_8) \bigg] t^A\psi \bigg)
  \bigg\}.
\end{align}
We rewrite the Hamiltonian in terms of the \textit{effective}
four-quark operators that contain already the Wilson coefficients
$\mathcal{Q}_i(z;\mu)$
\begin{align}
\mathcal{H}_{W}^{\Delta I=1} =
  \frac{G_F}{\sqrt{6}}\frac{\sin^2\theta_W}{3}\left(
\mathcal{Q}_1+\mathcal{Q}_2+\mathcal{Q}_3+\mathcal{Q}_4\right),
\end{align}
where the four-quark operators $\mathcal{Q}_i(z;\mu)$ are defined as
\begin{align}
\mathcal{Q}_i(z;\mu) = \alpha_i \left( \bar{\psi} \Gamma^{(i)}_1 \psi \right)
  \left( \bar{\psi} \Gamma^{(i)}_2 \psi \right),
\end{align}
where $\alpha_i=1$ for $i=1,\, 3$ and $\alpha_i = 1/2$ for $i=2,\, 4$.
The $\Gamma^{(i)}_j$ are defined as
\begin{align}
\Gamma_1^{(1)}&=\gamma_\mu\gamma_5\lambda_3, \;\;
                \Gamma_2^{(1)}=\gamma_\mu\Lambda^{(1)}, \;\;\;\;\;\;\;
\Gamma_1^{(2)}=\gamma_\mu\gamma_5\lambda_3 t^A, \;\;
                \Gamma_2^{(2)}=\gamma_\mu\Lambda^{(2)}t^A, \cr
\Gamma_1^{(3)}&=\gamma_\mu\lambda_3, \;\;\;\;\;
                \Gamma_2^{(3)}=\gamma_\mu\gamma_5\Lambda^{(3)},
                \;\;\;\; \Gamma_1^{(4)}=\gamma_\mu\lambda_3t^A,
                \;\;\;\;\;\,
                \Gamma_2^{(4)}=\gamma_\mu\gamma_5\Lambda^{(4)}t^A
\label{eq:Gamma}
\end{align}
with flavor matrices defined as
\begin{align}
\Lambda^{(1)} &= \lambda_0\bigg(\sqrt{2}c_1+\frac{c_5}{\sqrt{2}}\bigg)
  +\lambda_8(c_1-c_5), \;\;\;\;
\Lambda^{(2)}=\lambda_0
  \bigg(\sqrt{2}c_2+\frac{c_6}{\sqrt{2}}\bigg)+\lambda_8(c_2-c_6), \cr
\Lambda^{(3)}& = \lambda_0
               \bigg(\sqrt{2}c_3+\frac{c_7}{\sqrt{2}}\bigg) +
               \lambda_8(c_3-c_7),\;\;\;\;
\Lambda^{(4)} = \lambda_0 \bigg(\sqrt{2}c_4 +
               \frac{c_8}{\sqrt{2}}\bigg) + \lambda_8(c_4-c_8)\ .
\end{align}

In order to compute the $\Delta I=1$ flavor-conserving effective weak
chiral Lagrangian, we employ the N$\chi$QM from the instanton
vacuum. The effective weak chiral Lagrangian is defined as a
vacuum expectation value (VEV) of the effective weak
Hamiltonian~\cite{Franz:1999wr, Franz:1999ik}
\begin{align}
 \mathcal{L}_W^{\Delta I=1} = \int D \psi D
  \psi^\dagger  \mathcal{H}_W^{\Delta I=1} \exp \left[
\int d^4 z\,\psi^\dagger(z) \mathcal{D} \psi(z) \right],
\end{align}
where $\mathcal{D}$ represents the nonlocal covariant Dirac operator
defined as
\begin{align}
\mathcal{D}(-i\partial) \equiv i \gamma_\mu \partial_\mu +
  i\sqrt{M(-i\partial)} U^{\gamma_5}(x) \sqrt{M(-i\partial)},
\end{align}
where $U^{\gamma_5}$ represents the chiral field defined as
\begin{equation}
U^{\gamma_5} \;=\; \frac{1+\gamma_5}{2} U + \frac{1-\gamma_5}{2}
U^\dagger
\end{equation}
with the Goldstone boson field $U=\exp(i\lambda^a \pi^a/f_\pi)$.
Then, the flavor-conserving effective weak chiral Lagrangian can be
expressed in terms of the VEV of the four-quark operator
\begin{align}
{\cal L}_{\mathrm{eff}} = \frac{G_F}{\sqrt{6}} \;
  \frac{\sin^2\theta_W}{3} \sum_{i=1}^4\langle   \mathcal{Q}_i \rangle.
\end{align}
We refer to Refs.~\cite{Franz:1999wr, Franz:1999ik, Lee:2004tr} for
details of how to compute the VEV of $\mathcal{Q}_i$.

The flavor-conserving effective weak chiral Lagrangian in the $\Delta
I=1$ channel is obtained in terms of the low-energy constants
$\mathcal{N}_i$ and $\mathcal{M}_i$
\begin{align}
\mathcal{L}_{\mathrm{eff}}^{\Delta I=1} & =
\mathcal{N}_1\langle(R_\mu-L_\mu)\lambda_3  \rangle
\langle(R_\mu+L_\mu)\lambda_0\rangle
+ \mathcal{N}_2\langle(R_\mu-L_\mu)\lambda_3\rangle
\langle(R_\mu+L_\mu)\lambda_8\rangle
\cr
&+\mathcal{N}_3\langle(R_\mu-L_\mu)\lambda_0\rangle
\langle(R_\mu+L_\mu)\lambda_3\rangle
+\mathcal{N}_4 \langle(R_\mu-L_\mu)\lambda_8\rangle
\langle(R_\mu+L_\mu)\lambda_3\rangle
\cr
& + \mathcal{N}_{5} \langle \lambda_3 U\lambda_0U^\dagger - \lambda_3
  U^\dagger\lambda_0U \rangle
+ \mathcal{N}_{6} \langle \lambda_3
  U\lambda_8U^\dagger-\lambda_3 U^\dagger\lambda_8U \rangle
\cr
& + \mathcal{N}_{7} \langle L_\mu\lambda_3 L_\mu U^\dagger\lambda_0U
-R_\mu\lambda_3 R_\mu U\lambda_0 U^\dagger \rangle
\cr
& + \mathcal{N}_{8} \langle L_\mu\lambda_3 L_\mu U^\dagger\lambda_8U
-R_\mu\lambda_3 R_\mu U\lambda_8U^\dagger \rangle
\cr
& + \mathcal{N}_{9} \langle \left(\lambda_3 R_\mu R_\mu
+ R_\mu R_\mu\lambda_3\right) U\lambda_0U^\dagger
-\left(\lambda_3 L_\mu L_\mu
+L_\mu L_\mu \lambda_3\right) U^\dagger\lambda_0U \rangle
\cr
& + \mathcal{N}_{10} \langle \left(\lambda_3 R_\mu R_\mu
+ R_\mu R_\mu\lambda_3 \right) U\lambda_8U^\dagger
-\left(\lambda_3 L_\mu L_\mu
+L_\mu L_\mu \lambda_3 \right) U^\dagger\lambda_8U \rangle
\cr
& + \mathcal{N}_{11} \langle (R_\mu\lambda_3 R_\mu-L_\mu\lambda_3
  L_\mu) \lambda_0\rangle
+ \mathcal{N}_{12} \langle (R_\mu\lambda_3 R_\mu-L_\mu\lambda_3 L_\mu)
  \lambda_8\rangle,
\cr
& + \mathcal{M}_1 \langle \lambda_3 U \lambda_0U^\dagger - \lambda_3
  U^\dagger \lambda_0U \rangle
+ \mathcal{M}_2 \langle \lambda_3 U\lambda_8U^\dagger - \lambda_3
  U^\dagger \lambda_8U \rangle
\cr
&+ \mathcal{M}_3 \langle L_\mu\lambda_3 L_\mu U^\dagger\lambda_0U
-R_\mu\lambda_3 R_\mu U\lambda_0U^\dagger\rangle
+ \mathcal{M}_4 \langle L_\mu\lambda_3 L_\mu U^\dagger\lambda_8U
-R_\mu\lambda_3 R_\mu U\lambda_8U^\dagger\rangle
\cr
& + \mathcal{M}_5 \langle \left(\lambda_3 R_\mu R_\mu
+ R_\mu R_\mu\lambda_3 \right) U\lambda_0U^\dagger
-\left( \lambda_3 L_\mu L_\mu
+L_\mu L_\mu \lambda_3 \right) U^\dagger\lambda_0U \rangle
\cr
& + \mathcal{M}_{6} \langle \left(\lambda_3 R_\mu R_\mu
+ R_\mu R_\mu\lambda_3 \right) U\lambda_8U^\dagger
-\left(\lambda_3 L_\mu L_\mu
+L_\mu L_\mu \lambda_3 \right) U^\dagger\lambda_8U \rangle
\cr
& + \mathcal{M}_{7} \langle (R_\mu\lambda_3 R_\mu-L_\mu\lambda_3
  L_\mu) \lambda_0 \rangle
+ \mathcal{M}_{8} \langle (R_\mu\lambda_3 R_\mu-L_\mu\lambda_3 L_\mu)
  \lambda_8 \rangle
\label{eq:ExL1}
\end{align}
where $\langle \cdots \rangle$ means the trace over the flavor. The
right and left currents $R_\mu$ and $L_\mu$ are defined respectively
as
\begin{align}
R_\mu = i U \partial_\mu U^\dagger,\,\,
L_\mu = i U^\dagger \partial_\mu U\ .
\end{align}
The weak low-energy constants (WLECs) $\mathcal{N}_i$ are the leading
order in the large $N_c$ limit whereas $\mathcal{M}_i$ are of the
subleading order. They are expressed as
\begin{align}
\mathcal{N}_1 &= 4 N_c^2 J_2^2 \mathcal{C} \left(
                \sqrt{2}c_1+\frac{c_5}{\sqrt{2}}\right),\;\;\;\;
\mathcal{N}_2 = 4 N_c^2 J_2^2 \mathcal{C}\left(c_1-c_5 \right),\cr
\mathcal{N}_3 & = 4 N_c^2 J_2^2 \mathcal{C} \left(\sqrt{2}c_3 +
                \frac{c_7}{\sqrt{2}}\right), \;\;\;\;
\mathcal{N}_4 = 4 N_c^2 J_2^2 \mathcal{C} \left(c_3-c_7\right), \
\cr
\mathcal{N}_5 & = 8 N_c^2 J_1^2 \mathcal{C} \left(\sqrt{2}c_2 - \sqrt{2}c_4 +
                \frac{c_6}{\sqrt{2}}-\frac{c_8}{\sqrt{2}}\right),\;\;\;\;
\mathcal{N}_6  = 8 N_c^2 J_1^2 \mathcal{C}\left(c_2-c_4-c_6+c_8\right),
\cr
\mathcal{N}_7 & = 16 N_c^2 J_1 J_3 \mathcal{C} \left(\sqrt{2}c_2 -
                \sqrt{2}c_4 + \frac{c_6}{\sqrt{2}}-\frac{c_8}{\sqrt{2}}\right),\;\;\;\;
\mathcal{N}_8  = 16 N_c^2 J_1 J_3
                \mathcal{C}\left(c_2-c_4-c_6+c_8\right),
\cr
\mathcal{N}_9 & = 8 N_c^2 J_1J_4 \mathcal{C} \left(\sqrt{2}c_2 - \sqrt{2}c_4 +
                \frac{c_6}{\sqrt{2}}-\frac{c_8}{\sqrt{2}}\right),\;\;\;\;
\mathcal{N}_{10}  = 8 N_c^2 J_1 J_4
                \mathcal{C}\left(c_2-c_4-c_6+c_8\right),
\cr
\mathcal{N}_{11} & = 4 N_c^2 J_2^2  \mathcal{C}\left(\sqrt{2}c_2 +
                   \sqrt{2}c_4 +
                   \frac{c_6}{\sqrt{2}}+\frac{c_8}{\sqrt{2}}\right),
                   \;\;\;\;
\mathcal{N}_{12} = 4 N_c^2 J_2^2  \mathcal{C}
                   \left(c_2+c_4-c_6-c_8\right), \cr
\mathcal{M}_1 & = 8 N_c J_1^2 \mathcal{C} \left(
                \sqrt{2}c_1 - \sqrt{2}c_3 +
                \frac{c_5}{\sqrt{2}}-\frac{c_7}{\sqrt{2}}\right), \;\;\;\;
\mathcal{M}_2  = 8 N_c J_1^2 \mathcal{C}\left(c_1-c_3-c_5+c_7\right),
\cr
\mathcal{M}_3 & = 16 N_c J_1 J_3 \mathcal{C} \left(
                \sqrt{2}c_1 - \sqrt{2}c_3 +
                \frac{c_5}{\sqrt{2}}-\frac{c_7}{\sqrt{2}}\right), \;\;\;\;
\mathcal{M}_4  = 16 N_c J_1 J_3 \mathcal{C}\left(c_1-c_3-c_5+c_7\right),
\cr
\mathcal{M}_5 & = 8 N_c J_1 J_4 \mathcal{C} \left(
                \sqrt{2}c_1 - \sqrt{2}c_3 +
                \frac{c_5}{\sqrt{2}}-\frac{c_7}{\sqrt{2}}\right), \;\;\;\;
\mathcal{M}_6  = 8 N_c J_1 J_4 \mathcal{C}\left(c_1-c_3-c_5+c_7\right),
\cr
\mathcal{M}_7 & = 4 N_c J_2^2 \mathcal{C} \left( \sqrt{2}c_1 +
  \sqrt{2}c_3 + \frac{c_5}{\sqrt{2}}+\frac{c_7}{\sqrt{2}}\right),\;\;\;\;
\mathcal{M}_8  = 4 N_c J_2^2 \mathcal{C} \left(c_1+c_3-c_5-c_7\right),
\label{eq:LECs}
\end{align}
where the integrals $J_1$, $J_2$, $J_3$, and $J_4$ are defined
respectively as
\begin{align}
J_{1}&=-\int \frac{d^4 k}{(2\pi)^4}\ \frac{M(k)}{k^2+M^2(k)}
=\frac{\left\langle\overline{\psi}{\psi}\right\rangle_{M}}{4N_c},
\cr
J_{2}&=\int \frac{d^4 k}{(2\pi)^4}\ \frac{M^2(k)
-k^2M(k) \tilde{M}^\prime + k^4 M'^2(k)}{(k^2+M^2(k))^2} =
       \frac{f_\pi^2}{4N_c},
\cr
J_{3}&=\int \frac{d^4 k}{(2\pi)^4}\
\left[\frac{\frac{1}{4}\tilde{M}^{\prime\prime}k^2
+\frac{1}{2}\tilde{M}^\prime-\frac{\tilde{M}^{\prime2}}{8M}k^2}{k^2+M^2(k)}
       \right.        \cr
&-\left. \frac{M+M^2\tilde{M}^\prime +
  \frac{k^2}{2}M^2\tilde{M}^{\prime\prime} +
  \frac{1}{2}k^2M\tilde{M}^{\prime2}
+\frac{k^2}{4}\tilde{M}^\prime}{(k^2+M^2(k))^2}
+k^2\frac{\frac{1}{2}M+2M^2\tilde{M}^\prime
+M^3\tilde{M}^{\prime2}}{(k^2+M^2(k))^3}\right],\cr
J_4&=\int \frac{d^4 k}{(2\pi)^4}\
\frac{-M^3+k^2M^2\tilde{M}^\prime}{(k^2+M^2(k))^3}.
\label{eq:coefficient4}
\end{align}
$M(k)$ represents the momentum-dependent quark mass, and $\tilde{M}'$
and $\tilde{M}''$ are defined as
\begin{align}
\tilde{M}'\equiv \frac{M'}{2|k|}\ ,\ \
\tilde{M}''\equiv \frac{1}{4|k|^3}(M''|k|-M') .
\end{align}
The $\mathcal{C}$ contains the Fermi constant and the Weinberg angle
\begin{align}
\mathcal{C} = \frac{G_F}{\sqrt{6}} \frac{\sin^2 \theta_W}{3}.
\end{align}

To compute the WLECs in Eq.(\ref{eq:LECs}), we use the
momentum-dependent quark mass derived from the instanton
vacuum~\cite{Diakonov:1985eg}
\begin{align}
M(k) = M_0 F^2(k\rho)
\end{align}
with
\begin{align}
F(k\rho) = 2 z \left( I_0(z)K_1(z) - I_1(z) K_0(z) - \frac{1}{z}
  I_1(z) K_1(z)  \right),
\end{align}
where $I_i$ and $K_i$ are the modified Bessel functions, and 
$z =k/2\Lambda$. 
The value of the
dynamical quark mass at the zero virtuality of the quark is also
obtained from the instanton vacuum, i.e. $M_0=350\,\mathrm{MeV}$,
given the average size of the instanton and the interdistance between
instantons $R\approx 1\,\mathrm{fm}$~\cite{Diakonov:1985eg}.
The parameter $\Lambda$ is determined to reproduce the physical
value of $f_\pi$ through Eq.~(\ref{eq:coefficient4}).

As discussed already in Ref.~\cite{Franz:1999ik}, the vector and
axial-vector currents are not conserved in the presence of the
nonlocal interaction arising from the momentum-dependent quark
mass, that is, the corresponding gauge symmetries are broken. In order
to keep the currents conserved, we need to make the effective chiral
action gauge-invariant. In Ref.~\cite{Musakhanov:2002xa, Kim:2004hd},
the gauged effective chiral action was derived, based on the instanton
vacuum.  Had we naively computed $J_2=f_\pi^2/4N_c$ without the current
conservation being considered, then we would have ended up with
the Pagels-Stokar formula for $f_\pi^2$~\cite{Pagels:1979hd}, which
does not satisfy the gauge invariance. The numerical
results for the integrals given in Eq.(\ref{eq:coefficient4}) are
obtained as
\begin{align}
J_1=(-112.31)^3\,\mathrm{MeV}^3, \;\;\;\;
J_2=(26.673)^2\,\mathrm{MeV}^2,\;\;\;\;
J_3=-1.7403\,\mathrm{MeV},\;\;\;\;
J_4=-0.601\,\mathrm{MeV}.
\label{eq:integrals}
\end{align}
Note that the value of $J_1$ is related to that of the quark
condensate $\langle \overline{\psi}\psi\rangle_M
=(-257.13\,\mathrm{MeV})^3$ and that of $J_2$ corresponds to the value
of $f_\pi=92.4$ MeV.

\begin{table}[htp]
\centering
\begin{tabular}{ccccc} \hline\hline
  &   LO & NLO ($Z$) & NLO ($Z+W$) &KS  \\ \hline
$c_1$ &  $\phantom{-}0.264$ & $-0.054$ & $-0.055$ & $\phantom{-}0.403$
  \\
$c_2$ &  $\phantom{-}0.981$ & $\phantom{-}0.803$ & $\phantom{-}0.810$
                                &  $\phantom{-}0.765$ \\
$c_3$ &  $-0.592$ & $-0.629$ & $-0.627$ & $-0.463$  \\
$c_4$ &  $0$ & $0$ & $0$ & $0$ \\ \hline
$c_5$ &  $\phantom{-}5.97$ & $\phantom{-}4.85$ & $\phantom{-}5.09$ &
             $\phantom{-}5.61$ \\
$c_6$ &  $-2.30$ & $-2.14$ & $-2.55$ & $-1.90$ \\
$c_7$ &  $\phantom{-}5.12$ & $\phantom{-}4.27$ & $\phantom{-}4.51$ &
         $\phantom{-}4.74$   \\
$c_8$ &  $-3.29$ & $-2.94$ & $-3.36$ & $-2.67$  \\
\hline \hline
\end{tabular}
\caption{The Wilson coefficients $c_i$ derived from
  Ref.~\cite{Tiburzi:2012hx}. The last column denoted by KS lists the
  values derived in Ref.~\cite{ks1993} at the one-loop level.}
\label{tab:cs}
\end{table}
In Table~\ref{tab:cs}, the values of the Wilson coefficients are
listed. Those in the first three columns are taken from
Ref.~\cite{Tiburzi:2012hx}. The first column lists the results for the
Wilson coefficients in the LO, whereas the second and third ones
correspond to those from the NLO contributions together with the
LO terms. The $Z$ and $Z+W$ in the second and
third columns stand respectively for the considerations of $Z$ and
$Z+W$ boson exchanges. The last column presents those at the one-loop
level from Ref.~\cite{ks1993}. As already discussed in
Ref.~\cite{Tiburzi:2012hx}, there are certain effects from the NLO
contributions.

\begin{table}[htp]
\centering
\begin{tabular}{ccccc} \hline\hline
WLEC & LO & NLO($Z$) & NLO($Z+W$) & KS  \\ \hline
${\cal N}_1$ & $\phantom{-}0.307$ & $\phantom{-}0.224$ &
               $\phantom{-}0.235$ & $\phantom{-}0.303$ \\
${\cal N}_2$ & $-0.381$ & $-0.328$ & $-0.344$ & $-0.348$ \\
${\cal N}_3$ & $\phantom{-}0.186$ & $\phantom{-}0.142$ &
               $\phantom{-}0.154$ & $\phantom{-}0.180$ \\
${\cal N}_4$ & $-0.382$ & $-0.327$ & $-0.343$ & $-0.348$ \\
${\cal N}_5$ & $\phantom{-}1.106$ & $\phantom{-}0.901$ &
               $\phantom{-} 0.910$ & $\phantom{-}0.861$ \\
${\cal N}_6$ & $-0.005$ & $\phantom{-} 0.002$ & $0$ & $-0.003$ \\
${\cal N}_7$ & $\phantom{-} 2.716$ & $\phantom{-} 2.214$ &
               $\phantom{-} 2.236$ & $\phantom{-} 2.117$ \\
${\cal N}_8$ & $-0.012$ & $\phantom{-}0.004$ & $0$ & $-0.007$ \\
${\cal N}_9$ & $\phantom{-} 0.469$ & $\phantom{-} 0.382$ &
               $\phantom{-} 0.386$ & $\phantom{-} 0.365$ \\
${\cal N}_{10}$ & $-0.002$ & $\phantom{-}0.001$ & $0$ & $-0.001$ \\
${\cal N}_{11}$ & $-0.171$ & $-0.164$ & $-0.203$ & $-0.144$ \\
${\cal N}_{12}$ & $\phantom{-} 0.439$ & $\phantom{-} 0.393$ &
                 $\phantom{-} 0.449$ & $\phantom{-} 0.356$
\\ \hline
${\cal M}_1$ & $\phantom{-}0.320$ & $\phantom{-}0.216$ &
               $\phantom{-}0.215$ & $\phantom{-} 0.325$ \\
${\cal M}_2$ & $\phantom{-}0.001$ & $-0.001$ & $-0.001$ & $-0.001$  \\
${\cal M}_3$ & $\phantom{-}0.786$ & $\phantom{-}0.531$ &
               $\phantom{-} 0.529$ & $\phantom{-} 0.798$ \\
${\cal M}_4$ & $\phantom{-}0.003$ & $-0.002$ & $-0.004$ & $-0.002$ \\
${\cal M}_5$ & $\phantom{-}0.136$ & $\phantom{-}0.092$ &
               $\phantom{-}0.091$ & $\phantom{-}0.138$ \\
${\cal M}_6$ & $\phantom{-}0.0004$ & $-0.0004$ & $-0.001$ & $-0.0003$ \\
${\cal M}_7$ & $\phantom{-}0.164$ & $\phantom{-}0.122$ &
               $\phantom{-}0.130$ & $\phantom{-}0.161$ \\
${\cal M}_8$ & $-0.254$ & $-0.218$ & $-0.229$ & $-0.232$ \\
\hline\hline
\end{tabular}
\caption{The results for the low-energy constants in units of in
  $10^{-4}~{\rm MeV}^2$. Notations are the same as in
  Table~\ref{tab:cs}.}
\label{tab:lecs}
\end{table}
Based on these values of the Wilson coefficients, we
list in Table~\ref{tab:lecs} the results for the WLECs given in
Eq.(\ref{eq:LECs}).  Note that the WLECS ${\cal N}_6$, ${\cal N}_8$, and ${\cal N}_{10}$
are null. This is due to the fact that they correspond to the
operators
\begin{align}
O_2'= -O_2 + O_4 + O_6 - O_8,
\end{align}
for which the corresponding Wilson coefficient vanishes because
$O_2'$ is not generated by QCD radiative
corrections~\cite{Tiburzi:2012hx}.

Though there are arguments that sizable contributions in $\Delta I=1$
channel come from the operators with strangeness~\cite{ks1993}, we
will restrict ourselves to the case of SU(2). The calculation in SU(2)
has several merits in particular in the present work. Firstly, the
chiral solitonic approach in SU(2) is much simpler and physically
clearer than that in SU(3). Secondly, the SU(2) approach allows one to 
understand better the PV $\pi NN$ constant based on the effective weak
Hamiltonian. A more quantitative work within SU(3) will appear
elsewhere. In the case of SU(2), we reduce $\lambda_0$, $\lambda_3$, and
$\lambda_8$  to
\begin{align}
\lambda_0~\rightarrow~\sqrt{\frac{2}{3}}\bm{1},\;\;\;\;
\lambda_3~\rightarrow~\tau_3,\;\;\;\;
\lambda_8~\rightarrow~\frac{1}{\sqrt{3}}\bm{1}.
\end{align}
As a result, the $\Delta I=1$ effective weak Lagrangian is simplified
as
\begin{align}
\mathcal{L}_{\mathrm{eff}}^{\mathrm{SU(2)}}
&= \beta_1 \langle(R_\mu-L_\mu)\tau_3\rangle \langle
  R_\mu+L_\mu\rangle
 + \beta_2
\langle R_\mu-L_\mu\rangle \langle(R_\mu+L_\mu)\tau_3\rangle
\cr
& + \beta_3
\langle (R_\mu\ R_\mu-L_\mu L_\mu)\tau_3\rangle,
+ \beta_4 \langle (R_\mu\ R_\mu-L_\mu L_\mu)\tau_3\rangle,
\label{eq:su2Lag}
\end{align}
where $\beta_i$ are defined in terms of the WLECs
\begin{align}
\beta_1 &= \frac1{\sqrt{3}}\left(\sqrt{2} \mathcal{N}_1 +
           \mathcal{N}_2\right), \;\;\;\;
\beta_2 = \frac1{\sqrt{3}}\left(\sqrt{2} \mathcal{N}_3 +
           \mathcal{N}_4\right),\cr
\beta_3 &= \frac1{\sqrt{3}}\left[\sqrt{2} \mathcal{N}_{11} + \mathcal{N}_{12}
  +  \sqrt{2}(2\mathcal{N}_9 -\mathcal{N}_7) +
 2 \mathcal{N}_{10} - \mathcal{N}_8  \right],\cr
\beta_4 &= \frac{1}{\sqrt{3}}\left[\sqrt{2}\mathcal{M}_7 + \mathcal{M}_8
  + \sqrt{2}(2\mathcal{M}_5 -\mathcal{M}_3) +
  2\mathcal{M}_{6} - \mathcal{M}_4 \right].
\label{eq:betas}
\end{align}
As will be shown soon, $\beta_1$ and $\beta_2$ do not
contribute at all to the PV $\pi NN$ coupling constant. On the other
hand, $\beta_3$ and $\beta_4$ do come into play, so that we need to
examine them in detail. We can explicitly express $\beta_3$ and
$\beta_4$ in terms of the Wilson coefficients such that we can see
which terms contribute dominantly to the PV $\pi NN$ coupling
constant. $\beta_3$ and $\beta_4$ are rewritten as
\begin{align}
\beta_3 &= \frac{G_F\sin^2 \theta_W }{12\sqrt{2}} \left[
(c_2+c_4) f_\pi^4 - 16 N_c (c_2-c_4) \langle \overline{\psi}\psi\rangle_M
          (J_3-J_4)  \right],\cr
\beta_4 &=\frac{G_F\sin^2 \theta_W }{12\sqrt{2}N_c} \left[
(c_1+c_3) f_\pi^4 - 16 N_c (c_1-c_3) \langle \overline{\psi}\psi\rangle_M
          (J_3-J_4)  \right],
  \label{eq:beta34}
\end{align}
which clearly shows that $\beta_4$ is the subleading order in the large
$N_c$ limit with respect to $\beta_3$. Note that the structure of the
$\beta_4$ is the same as that of $\beta_3$ except for the Wilson
coefficients and the $1/N_c$ factor. The magnitudes of the second
terms in Eq.(\ref{eq:beta34}) are much larger than those of the first
ones, we can ignore approximately the first terms. That is, $\beta_3$
and $\beta_4$ can be expressed as
\begin{align}
\beta_3 \approx  \frac{G_F\sin^2 \theta_W }{12\sqrt{2}} c_2 \langle
  \overline{\psi}\psi \rangle_M (J_4-J_3),\;\;\;\;
\beta_4 \approx -\frac{c_3}{c_2 N_c} \beta_3,
\label{eq:betaapprx}
\end{align}
which indicates that $\beta_3$ is larger than $\beta_4$ approximately
by $(70-75)\,\%$.

\begin{table}[htp]
\centering
\begin{tabular}{ccccc} \hline \hline
WLEC & LO & NLO($Z$) & NLO($Z+W$) & KS  \\ \hline
$\beta_1$ & $\phantom{-}0.031$ & $-0.006$ & $-0.006$ &
            $\phantom{-}0.047$ \\
$\beta_2$ & $-0.069$ & $-0.073$ & $-0.073$ & $-0.054$ \\
$\beta_3$ & $-1.334$ & $-1.092$ & $-1.102$ & $-1.041$ \\
$\beta_4$ & $-0.434$ & $-0.309$ & $-0.308$ & $-0.428$ \\
\hline\hline
\end{tabular}
\label{tab:betas}
\caption{The values of $\beta_i$ in units of $10^{-4}~{\rm
    MeV}^2$. Notations are the same as in Table~\ref{tab:cs}.}
\end{table}
In Table~\ref{tab:betas}, we list the results for the $\beta_i$. Note
that $\beta_3$ and $\beta_4$ have the same sign because $c_2$ and
$c_3$ have different relative signs as shown in Table~\ref{tab:cs}.
The magnitude of $\beta_3$ indeed turns out to be about 75 \% larger
than the $\beta_4$, as expected from Eq.(\ref{eq:betaapprx}).

\section{Parity-violating $\pi NN$ coupling
  constant}\label{sec:result}
We are now in a position to determine the PV $\pi NN$ coupling
constant. Starting from Eq.(\ref{eq:su2Lag}), we are able to derive
the PV $\pi NN$ coupling constant. We already have shown explicitly
how one can obtain the PV $\pi NN$ coupling constant, based on the
$\chi$QSM~\cite{Lee:2012fx}. Thus, we want to briefly explain the
procedure of computing the $h_{\pi NN}^1$ within the model. The PV
$\pi NN$ coupling constant can be derived by solving the following
matrix element:
\begin{align}
\langle N| \mathcal{H}_W^{\Delta I=1} |\pi^a N\rangle  &=
 \frac{G_F}{\sqrt{6}} \frac{\sin^2 \theta_W}{3} \sum_{i=1}^4
 \langle N | \mathcal{Q}_i(z;\mu) |\pi^a N\rangle  \cr
&= \frac{G_F}{\sqrt{6}} \frac{\sin^2 \theta_W}{3} \sum_{i=1}^4 \int
  d^4 \xi    (k^2 + m_\pi^2) e^{ik\cdot\xi} \langle N |
  \mathcal{T} [\mathcal{Q}_i(z;\mu) \pi^a(\xi)]|N\rangle,
\label{eq:mat1}
\end{align}
where the nucleon states can be constructed by using the Ioffe-type
current in Euclidean space ($x_{0}=-ix_{4}$)~\cite{Diakonov:1987ty,
  Christov:1995vm}:
\begin{align}
|N(p_{1})\rangle &= \lim _{y_{4}\rightarrow -\infty }
e^{p_{4}y_{4}}{\mathcal{N}}^{*}(p_{1})\int d^{3}y
e^{i{\bm p}_{1}\cdot {\bm y}}J^{\dagger }_{N}(y)|0\rangle,\cr
\langle N(p_{2})| &= \lim _{x_{4}\rightarrow +\infty }
e^{-p_{0}x_{4}}{\mathcal{N}}(p_{2})\int d^{3}x
e^{-i{\bm p}_{2}\cdot {\bm x}}\langle 0| J_{N}(x).
\end{align}
The  $J^{\dagger }_{N}$ ($J_{N}$) constitutes $N_{c}$ quarks
\begin{align}
J_{N}(x)\; =\; \frac{1}{N_{c}!}\epsilon ^{c_{1}c_{2}\cdots c_{N_{c}}}
\Gamma ^{s_{1} s_{2}\cdots
s_{N_{c}}}_{(TT_{3}Y)(JJ_{3}Y_{R})}\psi _{s_{1}c_{1}}(x)
\cdots \psi _{s_{N_{c}}c_{N_{c}}}(x)\,,
\end{align}
where $s _{1}\cdots s _{N_{c}}$ and $c_{1}\cdots c_{N_{c}}$
stand for spin-isospin and color indices, respectively.
The $\Gamma ^{\{s \}}_{(TT_{3})(JJ_{3})}$ provides
the quantum numbers $(TT_{3})(JJ_{3})$ for the nucleon: $T=1/2$, $Y=1$
and $J=1/2$. The nucleon creation operator $J_N^\dagger$ can be
obtained by taking the Hermitian conjugate of $J_N$.
The matrix elements in Eq.(\ref{eq:mat1}) is just the four-point
correlation function given as
\begin{align}
\label{Eq:correl}
\lim_{y_{0}\rightarrow -\infty \atop x_{0}\rightarrow +\infty }
\sum_{i=1}^{4} \langle 0 | {\mathcal
  T}[J_{N}(x)\mathcal{Q}^i(z;\mu) \partial_\mu A_\mu^a (\xi) J^{\dagger
}_{N}(y) ]|0\rangle\; =\;
\lim _{y_{0}\rightarrow -\infty
\atop x_{0}\rightarrow +\infty }{\mathcal{K}},
\end{align}
where $A_\mu^a$ stands for the axial-vector current. Note that we have
used the partial conservation of the axial-vector current
(PCAC). In principle, the four-point correlation function
$\mathcal{K}$ can be computed by solving the following functional
integral
\begin{equation}
\mathcal{K} \;=\; \frac1{\mathcal{Z}} \int D\psi D\psi^\dagger DU
J_{N}(x){\cal Q}^i(z;\mu) \partial_\mu A_\mu^a (\xi) J^{\dagger }_{N}(y)
\exp\left[\int d^4x \psi^\dagger \left(i\rlap{/}{\partial} + i
    \sqrt{M(-\partial^2)}U^{\gamma_5}\sqrt{M(-\partial)^2}\right)
  \psi\right].
\label{eq:corr2}
\end{equation}

As was already mentioned in the previous work~\cite{Lee:2012fx},
it is extremely complicated to deal with Eq.~(\ref{eq:corr2})
technically, since the PV $\pi NN$ coupling constant arises from both
the two-body quark operators $\mathcal{Q}^i$ and the axial-vector
one, which causes laborious triple sums over quark levels already at
the leading order in the large $N_c$ expansion. Thus, we employ the
gradient expansion method as in Ref.~\cite{Lee:2012fx}. In the
gradient expansion, $(\rlap{/}{\partial}U/M)\ll 1$ is used as an
expansion  method~\cite{Diakonov:1987ty} to expand the quark
propagator in the pion background field,  with the pion momentum
assumed to be small. Equivalently, we can directly start from the
effective weak chiral Lagrangian in
Eqs.(\ref{eq:ExL1},\ref{eq:su2Lag}) already derived in the previous
Section.

The classical soliton is assumed to have a hedgehog symmetry, so that
it can be parametrized in terms of the soliton profile function $P(r)$
\begin{align}
U_0 = \exp \left( i \bm{\tau} \cdot \hat{\bm{r}} \, P(r) \right).
\end{align}
In principle, $P(r)$ can be found by solving the equations of motion
self-consistently~\cite{Christov:1995vm}. However, we will employ a
parametrized form of $P(r)$ which is very close to the self-consistent
one.  The classical soliton field can be fluctuated such that the pion
field can be coupled to a $\Delta I=1$ two-body quark operator
\begin{align}
U=\exp\left(i\frac{\bm{\tau}\cdot \bm{\pi}}{2f_\pi}\right)U_0
\exp\left(i\frac{\bm{\tau}\cdot \bm{\pi}}{2f_\pi}\right).
\end{align}
Since the trace of the left and right currents over flavor space
vanish, i.e.
\begin{align}
\langle R_\mu+L_\mu\rangle=0,\;\;\;\;
\langle R_\mu-L_\mu\rangle=0,
\end{align}
the terms with $\mathcal{N}_1$, $\mathcal{N}_2$, $\mathcal{N}_3$, and
$\mathcal{N}_4$ do not contribute to $h^1_{\pi NN}$ as shown in our
previous analysis~\cite{Lee:2012fx} with the DDH effective
Hamiltonian~\cite{Desplanques:1979hn}. Considering the fact that
$\mathcal{N}_1$ and $\mathcal{N}_2$ contain the Wilson coefficient
$c_5$, which is the most dominant one, and $\mathcal{N}_3$ and
$\mathcal{N}_4$ have $c_7$ that is the second largest one, one can
explain a part of the reason why $h_{\pi NN}^1$ turns out to be rather
small in the present approach.

When it comes to all other terms, we can approximately rewrite
$L_\mu^2$ and $R_\mu^2$ as
\begin{align}
L_\mu L_\mu  \simeq
   \frac{i}{2f_\pi}\left(L_\mu^0L_\mu^0\bm{\tau}\cdot\bm{\pi}
-\bm{\tau}\cdot\bm{\pi}L_\mu^0L_\mu^0\right),\;\;\;\;
R_\mu R_\mu  \simeq \frac{i}{2f_\pi} \left(\bm{\tau}\cdot \bm{\pi}
              R_\mu^0R_\mu^0 -R_\mu^0R_\mu^0
              \bm{\tau}\cdot\bm{\pi}\right)
\end{align}
with $L^0_\mu=iU_0^\dagger\partial_\mu U_0$ and $R^0_\mu=iU_0\partial_\mu U_0^\dagger$,
so that we get
\begin{align}
\langle(R_\mu R_\mu -L_\mu L_\mu)\tau_3 \rangle
=i \frac{\sqrt{2}}{f_\pi}\langle(R_\mu^0 R_\mu^0+L_\mu^0L_\mu^0)
(\tau^-\pi^+ -\tau^+\pi^-)\rangle\ ,
\end{align}
where $\tau^\pm$ and $\pi^\pm$ are defined in the spherical basis as
\begin{align}
\tau^{\pm}=\mp \frac{1}{2}(\tau^1\pm i\tau^2)\ ,\
\pi^{\pm}=\mp \frac{1}{\sqrt{2}}(\pi^1\pm i\pi^2)\ .
\end{align}
The relevant effective Lagrangian is then expressed as
\begin{align}
\mathcal{L}_{\mathrm{eff}}^{\mathrm{SU(2)}}
=(\beta_3+\beta_4) \frac{i\sqrt{2}}{f_\pi}\langle(R_\mu^0
  R_\mu^0+L_\mu^0L_\mu^0) (\tau^-\pi^+ -\tau^+\pi^-)\rangle,
\end{align}
where $\beta_3$ and $\beta_4$ are defined already in
Eq.(\ref{eq:betas}).

Since we have already explained how the quantization of the soliton is
performed in the context of the PV $\pi NN$ coupling constant in
Ref.~\cite{Lee:2012fx}, we proceed to compute the $h_{\pi
  NN}^1$ within this framework. For simplicity, let us consider the PV
process  $n+\pi^+\to p$. Then, we need to compute the following trace
\begin{align}
\langle(R_\mu^0 R_\mu^0+L_\mu^0L_\mu^0)\tau^+\rangle .
\end{align}
Defining isovector fields $r^i_\mu$ and $l_\mu^i$ as
\begin{align}
R_\mu^0=-r_\mu^i\tau^i\ , \ L_\mu^0=-l_\mu^i\tau^i
\end{align}
and using an identity $\langle \tau^i\tau^j\tau^k\rangle =
2i\epsilon^{ijk}$, we obtain
\begin{align}
\langle R_\mu^0 R_\mu^0\tau^+\rangle =
(r_\mu^1+ir_\mu^2)r_\mu^3-r_\mu^3(r_\mu^1+ir_\mu^2),
\;\;\;\;
\langle L_\mu^0 L_\mu^0\tau^+\rangle
= (l_\mu^1+il_\mu^2)l_\mu^3-l_\mu^3(l_\mu^1+il_\mu^2).
\end{align}
Thus, we arrive at the final form of the effective Lagrangian
\begin{align}
{\cal L}^{\rm SU(2)}_{\rm eff}
=(\beta_3+\beta_4)  \frac{i\sqrt{2}}{f_\pi} \left[(r_\mu^1+ir_\mu^2)r_\mu^3
-  r_\mu^3(r_\mu^1+ir_\mu^2)+(r\rightarrow l)\right]\pi^+,
\end{align}
from which we can derive the PV $\pi NN$ coupling constant.
Using the collective quantization discussed in Ref.~\cite{Lee:2012fx},
we get
\begin{align}
&\int d^3x\langle p\uparrow|r_\mu^3(r_\mu^1+ir_\mu^2)|n\uparrow\rangle
=-\int d^3x\langle p\uparrow|(r_\mu^1+ir_\mu^2)r_\mu^3|n\uparrow\rangle
\cr
&=\frac{2\pi}{3I^2}\int dr\,r^2 \sin^2P(r)
  \left[\sin^2P(r)-3\cos^2P(r)\right],
\end{align}
where $I$ denotes the moment of inertia~\cite{Diakonov:1987ty}
expressed as
\begin{align}
I = \frac{N_c}{12} \int_{-\infty}^\infty
  \frac{d\omega}{2\pi}\mathrm{Tr}\left(\tau^i\frac1{\omega + i
  H}\tau^i\frac1{\omega + i  H} \right) \approx \frac83 \pi f_\pi^2
  \int_0^\infty dr\, r^2 \sin^2 P(r).
\end{align}
Here, $\omega$ is the energy frequencies of the quark levels and
$\mathrm{Tr}$ stands for the functional trace over coordinate space,
isospin and Dirac spin space. The second term was derived approximately
by the gradient expansion. Similarly, we obtain the same
result for $l_\mu$.  Having carried out the calculation of the matrix
element for the collective operators, we finally derive the PV $\pi
NN$ coupling constant  $h_{\pi NN}^1$ as
\begin{align}
h_{\pi NN}^1=i\langle p\uparrow|{\cal L}^{\rm SU(2)}_{\rm eff}|n\uparrow,
  \pi^+\rangle = \frac{8\sqrt{2}\pi}{3 f_\pi I^2}\, (\beta_3+\beta_4)\, \int dr\, r^2
  \sin^2P(r) \left[\sin^2P(r) - 3 \cos^2P(r)\right].
\label{eq:final}
\end{align}
It is interesting to see that Eq.(\ref{eq:final}) is exactly the same
as the expression obtained in Ref.~\cite{Lee:2012fx} except for the
coefficient $\beta_3+\beta_4$.

In order to compute the PV $\pi NN$ coupling constant, we
employ the following numerical values of the constants involved in the
present work: the Fermi constant $G_F=1.16637\times
10^{-5}\,\mathrm{GeV}^{-2}$, the Weinberg angle
$\sin^2\theta_W=0.23116$, and the pion decay constant is obtained to
be $f_\pi=0.0924\,\mathrm{GeV}$ given in
Eq.(\ref{eq:coefficient4}). Concerning the profile function,
we have already examined the dependence of $h_{\pi NN}^1$ on types of
the profile functions~\cite{Lee:2012fx}. The physical profile function
produces the largest value, compared to the linear and arctangent
profile functions. In the present work, we employ the physical profile
function expressed as
\begin{align}
P(r)=\left\{\begin{array}{ll}
2\arctan\left(\frac{r_0}{r}\right)^2, \hspace{2.0cm} \mbox{($r\leq r_x$)},\\
P_0 \, e^{-m_\pi r}(1+m_\pi r)/r^2, \hspace{0.7cm} \mbox{($r>r_x$)},\\
\end{array}\right.
\label{eq:profile}
\end{align}
where $r_0$ is defined as  $r_0=\sqrt\frac{3g_A}{16\pi f_\pi^2}$ with
the axial-vector constant $g_A=1.26$. $P_0$ and $r_x$ are given as
$P_0 =2r_0^2$, and $r_x=0.752\, \mathrm{fm}$, respectively. The
profile function in Eq.(\ref{eq:profile}) satisfies a correct behavior
of the Yukawa tail. Then, the moment of inertia is obtained to be
$I=3.32{\rm GeV}^{-1}$.

\begin{table}[htp]
\centering
\begin{tabular}{ccccc} \hline\hline
        & LO & NLO($Z$) & NLO($Z+W$) & KS  \\ \hline
$h_{\pi NN}^1$ & $10.96$ & $8.69$ & $8.74$ & $9.11$  \\
\hline\hline
\end{tabular}
\caption{$h_{\pi NN}^1$ in units of $10^{-8}$. Notations are the same
  as in Table~\ref{tab:cs}.}
\label{tab:h1pi}
\end{table}
Numerical results for $h_{\pi NN}^1$ are summarized in Table~\ref{tab:h1pi}.
In Ref.~\cite{Tiburzi:2012hx}, it was shown that NLO contributions
alter the values of the Wilson coefficients at $\mu=1$ GeV by about
$(10-20)\,\%$, which actually lessens the value of the $h_{\pi
  NN}^1$. by about $25\,\%$ as shown in Table~\ref{tab:h1pi}.
As already examined in Eqs.(\ref{eq:beta34}, \ref{eq:betaapprx}),
$\beta_3$ plays a dominant role in determining $h_{\pi NN}^1$. Thus,
the most important operator in the $\Delta I=1$ effective weak
Hamiltonian is $O_2$ in Eq.(\ref{eq:twobody}), which contains the
Wilson coefficient $c_2$. As clearly shown in Table~\ref{tab:h1pi},
the NLO QCD radiative corrections suppress the PV $\pi NN$ coupling
constant. In fact, we have shown already in the previous
work~\cite{Lee:2012fx}, the QCD radiative corrections strongly
diminish the value of $h_{\pi NN}^1$. This behavior contrasts
with the case of $K$ nonleptonic decays, where the penguin diagrams
enhance the contribution to the $\Delta I=1/2$ channel.
\begin{table}[htp]
\centering
\begin{tabular}{ccccccc} \hline\hline
DDH~\cite{Desplanques:1979hn} & DZ~\cite{dubovik1986}  &
KS~\cite{ks1993}  & QCD sum rules~\cite{henley1998}  & Skyrme
Model~\cite{weigel1999} & Lattice QCD~\cite{wasem2012}  & Present work
  \\ \hline
$45.6$ & $11.4$ & $60$ & $2$ & $8.0-1.3$ &
$10.99\pm 5.05_{-0.64}^{+0.58}$  &   $8.74$  \\
\hline\hline
\end{tabular}
\caption{Comparion of $h_{\pi NN}^1$ in units of $10^{-8}$ with
  various threoretical works. }
\label{tab:5}
\end{table}
In Table~\ref{tab:5}, we compare the present result with those of
various theoretical works. The present result turns out to be about 5
times smaller than the DDH ``best value''. We find that the result
from the QCD sum rules predicts the smallest value of $h^1_{\pi NN}$
whereas Ref.~\cite{ks1993} yields the largest result, in which the
importance of the strangeness contribution was emphasized. Compared
to the value of $h_{\pi NN}^1$ from lattice QCD with connected
diagrams considered only, the present result is in good agreement with
it.

\section{Summary and Outlook}\label{sec:con}
In the present work, we investigated the parity-violating
pion-nucleon coupling constant. Starting from the $\Delta I=1$
effective weak Hamiltonian~\cite{Tiburzi:2012hx} that considered
the next-to-leading order QCD radiative corrections, we derived the
effective weak chiral Lagrangian with the weak low-energy constants
determined in the $\Delta I=1$ and $\Delta S=0$ channel.
In order to calculate the parity-violating pion-nucleon coupling
constant $h_{\pi NN}^1$, we employed the chiral quark-soliton
model. Using the gradient expansion, which is equivalent to using the
effective weak chiral Lagrangian directly, we were able to compute
the values of $h_{\pi NN}^1$. We found that the first four terms of
the Lagrangian did not contribute at all to $h_{\pi NN}^1$, which
partially explains why the value of $h_{\pi NN}^1$ should be
small. It was also found that the main contribution to $h_{\pi NN}^1$
arose from the operator $O_2$ in the effective weak Hamiltonian.
We also noted that the next-to-leading-order QCD radiative corrections
further suppress the value of $h_{\pi NN}^1$ and as a result we
obtained $h_{\pi NN}^1=8.74\times 10^{-8}$. We compared this result with those from
various theoretical models including the recent result from lattice
QCD. The present result was shown to be in agreement with that
from lattice QCD.

The present work can be extended to the SU(3) case in which the
strange quark comes into play. Another merit of the chiral
quark-soliton model is that the explicit breaking of flavor
SU(3) symmetry can be treated systematically, the strange
quark mass being considered as a perturbation. Thus, it is interesting
to examine the contribution of the strange quark and its current quark
mass to the parity-violating pion-nucleon coupling constant. Other coupling
constants such as $h_{\rho NN}$ and $h_{\omega NN}$ can be studied
within the same framework. The related works are under way.

\section*{Acknowledments}
The work of H.-Ch.K. was supported by Basic Science
Research Program through the National Research Foundation of Korea
funded by the Ministry of Education, Science and Technology (Grant
Number: NRF-2015R1D1A1A01060707).
The work of H.J.L. was supported by Basic Science
Research Program through the National Research Foundation of Korea (NRF)
funded by the Ministry of Education (Grant
Number: NRF-2013R1A1A2009695).


\begin{thebibliography}{99}
\bibitem{RamseyMusolf:2006dz}
  M.~J.~Ramsey-Musolf and S.~A.~Page,
  Ann.\ Rev.\ Nucl.\ Part.\ Sci.\  {\bf 56} (2006) 1.

\bibitem{Holstein:2009zzc}
  B.~R.~Holstein,
  J.\ Phys.\ G {\bf 36} (2009) 104003.

\bibitem{Cirigliano:2013xha}
  V.~Cirigliano, S.~Gardner and B.~Holstein,
  Prog.\ Part.\ Nucl.\ Phys.\  {\bf 71} (2013) 93

\bibitem{Cirigliano:2013lpa}
  V.~Cirigliano and M.~J.~Ramsey-Musolf,
  Prog.\ Part.\ Nucl.\ Phys.\  {\bf 71} (2013) 2.

\bibitem{Haxton:2013aca}
  W.~C.~Haxton and B.~R.~Holstein,
  Prog.\ Part.\ Nucl.\ Phys.\  {\bf 71} (2013) 185.

\bibitem{Schindler:2013yua}
  M.~R.~Schindler and R.~P.~Springer,
  Prog.\ Part.\ Nucl.\ Phys.\  {\bf 72} (2013) 1.

\bibitem{Tadic:1969xx}
  D.~Tadic,
  Phys.\ Rev.\  {\bf 174} (1968) 1694.

\bibitem{Donoghue:1975xk}
  J.~F.~Donoghue,
  Phys.\ Rev.\ D {\bf 13} (1976) 2064.

\bibitem{Desplanques:1979hn}
  B.~Desplanques, J.~F.~Donoghue and B.~R.~Holstein,
  Annals Phys.\  {\bf 124} (1980) 449.

\bibitem{Zhu:2004vw}
  S.~L.~Zhu, C.~M.~Maekawa, B.~R.~Holstein, M.~J.~Ramsey-Musolf and U.~van Kolck,
  Nucl.\ Phys.\ A {\bf 748} (2005) 435.

\bibitem{haxton2013}
W. C. Haxton, and B. R. Holstein,
Prog. Part. Nucl. Phys. {\bf 71} (2013) 185.

\bibitem{schin2013}
M. R. Schindler, and R. P. Springer,
Prog. Part. Nucl. Phys. {\bf 72} (2013) 1.

\bibitem{hyun2001}
C. H. Hyun, T.-S. Park, and D.-P. Min,
Phys. Lett. B {\bf 516} (2001) 321.

\bibitem{hyun2005}
C. H. Hyun, S. J. Lee, J. Haidenbauer, and S. W. Hong,
Eur. Phys. J. A {\bf 24} (2005) 129.

\bibitem{hyun2007}
C. H. Hyun, S. Ando, and B. Desplanques,
Phys. Lett. B {\bf 651} (2007) 257.

\bibitem{des1986}
B. Desplanques, and J. J. Benayoun,
Nucl. Phys. A {\bf 458} (1986) 689.

\bibitem{kaiser1988}
N. Kaiser, and U. G. Meissner,
Nucl. Phys. A {\bf 489} (1988) 671.

\bibitem{weigel1999}
U. G. Meissner, and H. Weigel,
Phys. Lett. B {\bf 447} (1999) 1.

\bibitem{ks1993}
D. B. Kaplan, and M. J. Savage,
Nucl. Phys. A {\bf 556} (1993) 653.

\bibitem{dubovik1986}
V. M. Dubovik, and S. V. Zenkin,
Ann. Phys. {\bf 172} (1986) 100.

\bibitem{Lee:2004tr}
  H.~J.~Lee, C.~H.~Hyun, C.~H.~Lee and H.-Ch.~Kim,
  Eur.\ Phys.\ J.\ C {\bf 45} (2006) 451.

\bibitem{Lee:2012fx}
  H.~J.~Lee, C.~H.~Hyun and H.-Ch.~Kim,
  Phys.\ Lett.\ B {\bf 713} (2012) 439.

\bibitem{henley1998}
E. M. Henley, W. Y. P. Hwang, and L. S. Kisslinger,
Phys. Lett. B {\bf 367} (1996) 21.

\bibitem{wasem2012}
J. Wasem,
Phys. Rev. C {\bf 85} (2012) 022501.

\bibitem{Gericke:2011zz}
  M.~T.~Gericke {\it et al.},
  Phys.\ Rev.\ C {\bf 83}, 015505 (2011).


\bibitem{Korner:1978sz}
  J.~G.~Korner, G.~Kramer and J.~Willrodt,
  Phys.\ Lett.\ B {\bf 81} (1979) 365.

\bibitem{Guberina:1978wg}
  B.~Guberina, D.~Tadic and J.~Trampetic,
  Nucl.\ Phys.\ B {\bf 152} (1979) 429.

\bibitem{Karino:1980vn}
  T.~Karino, K.~Ohya and T.~Oka,
  Prog.\ Theor.\ Phys.\  {\bf 65} (1981) 693.

\bibitem{Karino:1981yj}
  T.~Karino, K.~Ohya and T.~Oka,
  Prog.\ Theor.\ Phys.\  {\bf 66} (1981) 1389.

\bibitem{Dai:1991bx}
  J.~Dai, M.~J.~Savage, J.~Liu and R.~P.~Springer,
  Phys.\ Lett.\ B {\bf 271} (1991) 403.

\bibitem{Tiburzi:2012hx}
  B.~C.~Tiburzi,
  Phys.\ Rev.\ D {\bf 85} (2012) 054020.

\bibitem{Franz:1999wr}
  M.~Franz, H.-Ch.~Kim and K.~Goeke,
  Nucl.\ Phys.\ B {\bf 562} (1999) 213.

\bibitem{Franz:1999ik}
  M.~Franz, H.-Ch.~Kim and K.~Goeke,
  Nucl.\ Phys.\ A {\bf 699} (2002) 541.

\bibitem{Diakonov:1985eg}
  D.~Diakonov and V.~Y.~Petrov,
  Nucl.\ Phys.\ B {\bf 272} (1986) 457.

\bibitem{Musakhanov:2002xa}
  M.~M.~Musakhanov and H.-Ch.~Kim,
  Phys.\ Lett.\ B {\bf 572} (2003) 181.

\bibitem{Kim:2004hd}
  H.-Ch.~Kim, M.~Musakhanov and M.~Siddikov,
  Phys.\ Lett.\ B {\bf 608} (2005) 95.

\bibitem{Pagels:1979hd}
  H.~Pagels and S.~Stokar,
  Phys.\ Rev.\ D {\bf 20} (1979) 2947.

\bibitem{Diakonov:1987ty}
  D.~Diakonov, V.~Y.~Petrov and P.~V.~Pobylitsa,
  Nucl.\ Phys.\ B {\bf 306} (1988) 809.

\bibitem{Christov:1995vm}
  C.~V.~Christov, A.~Blotz, H.-Ch.~Kim, P.~Pobylitsa, T.~Watabe,
  T.~Meissner, E.~Ruiz Arriola and K.~Goeke,
  Prog.\ Part.\ Nucl.\ Phys.\  {\bf 37} (1996) 91.
\end{thebibliography}
\end{document}